\documentclass[a4paper,fleqn,usenatbib]{mnras}
\usepackage{newtxtext,newtxmath}

\usepackage[T1]{fontenc}
\usepackage{ae,aecompl}

\usepackage{graphicx}	
\usepackage{amsmath}	
\usepackage{amssymb}	

\title[More from $V/V_m$]{Getting more out of $V/V_m$ than just  the mean}

\author[D. G. Banhatti]{
Dilip G. Banhatti,$^{1}$\thanks{E-mail: dilip.g.banhatti@gmail.com}
\\
$^{1}$Zeleni Trg 3A, 10000 Zagreb, Croatia\\
}

\date{Accepted XXX. Received YYY; in original form ZZZ}

\pubyear{2016}

\begin{document}
\label{firstpage}
\pagerange{\pageref{firstpage}--\pageref{lastpage}}
\maketitle

\begin{abstract} Banhatti earlier set down the procedure to derive cosmological number density $n(z)$ from the differential distribution $p(x)$ of the fractional luminosity volume relative to the maximum volume, $x \equiv V/V_m (0\leq x \leq 1)$, using a small sample of 76 quasars for illustrative purposes. This procedure is here applied to a bigger sample of 286 quasars selected from Parkes half-Jansky flat-spectrum survey at 2.7 GHz. The values of $n(z)$ are obtained for 8 values of redshift $z$ from 0 to 3.5. The function $n(z)$ can be interpreted in terms of redshift distribution obtained by integrating the radio luminosity function $\rho(P, z)$ over luminosities $P$ for the survey limiting flux density $S_0$ = 0.5 Jy.
\end{abstract}




\section{Introduction}

The luminosity-volume or $V/V_m$ test has traditionally been used only through the mean and standard error of $x \equiv V/V_m$ for cosmological samples of objects like quasars. Here $V$ is the volume upto the redshift $z$ of the object or quasar of flux density $S$ in a sample with limiting flux density $S_0$, and  $V_m$ the volume upto the limiting redshift $z_m$ at which the quasar would be as faint as possible and still be in the sample, i.e., have flux density $S_0$. For an unbiased sample, $x$ is calculated for each quasar and the mean and standard error are used to estimate uniformity in the space density of the underlying quasar population.

However, Kulkarni and Banhatti (1983) showed that the differential distributions $p(x)$ of $x$ and $n(z)$ of $z$ are directly related to each other. Banhatti (2009) illustrated how $n(z)$ is calculated from $p(x)$ for a small sample of 76 quasars. Here we use a bigger sample for a more realistic calculation. The cosmological models used in calculating $V$, $z_m$ and $V_m$ are only illustrative. Any other models could also be used.

Schmidt et al (1988) suggested use of $V/V_m$ test to examine the space distribution of gamma-ray bursts for homogeneity. The test was extensively used for a few years for this purpose (e.g., Dezalay et al 1994 and references therein), but the question of the location of gamma-ray bursts $\textendash$ whether within our Milky Way Galaxy or in other possibly distant galaxies $\textendash$ was not resolved from these studies. Only on discovering afterglows of gamma-ray bursts at lower (X-ray, optical, infrared and radio) photon energies in distant galaxies and thereafter measuring redshifts of the parent galaxies did it become clear that they form a cosmological population.

Use of $V/V_m$ distribution as outlined in this paper will be rewarding for a sufficiently large well-defined unbiased sample of gamma-ray bursts. It should be possible to construct such a sample from the results of space and ground-based gamma-ray telescopes by carefully taking into account detection methods and thresholds.



\section{Sample of quasars used and the world model}

Drinkwater et al (1997) define the survey and list the properties of 323 quasars from which 286 can be used for calculating $x \equiv V/V_m$. The sample used is thus 89\% complete relative to the survey, which covers 3.90 sr in the sky. Using the limiting flux density $S_0$ = 0.5 Jy at 2.7 GHz, the limiting redshift $z_m$ is calculated for each quasar from its redshift $z$, $\nu$ = 2.7 GHz flux density $S_\nu$, and spectral index $\alpha$ (defined by $\alpha \equiv - d(log S_\nu)/d(log \nu)$, or equivalently, $S_\nu \propto{}\nu^{-\alpha})$. The world model with the parameters $(q_0, \sigma_0, k, \lambda_0) = (1, 1, 1, 0)$ as defined by von Hoerner (1974) is used for the functions of $z$ needed, viz, the luminosity distance $\ell_\nu(z)$ and volume $v(z)$. These functions are:
\begin{equation}
(H_0/c)^2 \ell_\nu^2(\alpha, z) = z^2/(1+z)^{(1-\alpha)},
\end{equation}
\begin{equation}
(H_0/c)^3 v(z) = (3/2)(sin^{-1}f(z) - f(z)\surd[1-f(z)^2]),	
\end{equation}
where $f(z) = z/(1+z)$.

Here, $c/H_0 \equiv$ speed of light / Hubble constant, defines the linear scale.

\section{Deriving $\lowercase{n(z)}$ from $\lowercase{p}(V/V_{\lowercase{m}}) \equiv \lowercase{p(x)}$}
\subsection{Binning the $z_m$-values}
The quasars are first sorted out in increasing order of $z_m$. The limiting redshift is numerically calculated for each quasar using Newton-Raphson iteration (Rajarevathi 2007). The $z_m$-bins are then selected, so as to have roughly equal numbers of sources (about 30) each, which is good enough to derive the differential distribution of $x \equiv V/V_m (0 \leq x \leq 1)$ for each of the bins. Details of this binning are given in Table 1. Also listed are numbers proportional to the cosmological number densities $n(z_j)$ corresponding to the bin mid-points $z_j$. The procedure for calculating $n(z_j)$ is outlined later. Table 2 presents, for comparison, the same results for the smaller sample of 76 from Wills \& Lynds (1978) used by Banhatti (2009), although the world model used for those calculations is (von Hoerner (1974)) $(q_0, \sigma_0, k, \lambda_0) = (1/2, 1/2, 0, 0)$, for which the functions $\ell_\nu(z)$ and $v(z)$ are different (see Banhatti 2009).

\begin{table*}
	\centering
	\caption{Limiting redshifts, their bins, mid-points \& populations plus derived cosmological number densities using 286 quasars over 3.90 sr in the sky}
	\label{tab:Table 1}
	\begin{tabular}{lccccccccc} 
		\hline
		$z_m$-bin & 0 to 0.3 & 0.3 to 0.7 & 0.7 to 1.2 & 1.2 to 1.5 & 1.5 to 1.8 & 1.8 to 2.2 & 2.2 to 2.8 & 2.8 to 4.0 & > 4.0\\
		\hline
		$z_j$(bin mid-pt) & 0.15 & 0.5 & 0.95 & 1.35 & 1.65 & 2.0 & 2.5 & 3.4 & 300*\\
		\hline
		Bin pop. & 29 & 34 & 34 & 32 & 33 & 32 & 30 & 31 & 31\\
		\hline
		j (bin no.) & 1 & 2 & 3 & 4 & 5 & 6 & 7 & 8 & 9\\
		\hline
		$n(z_j)\propto$ & 48770. & 4717. & 1560. & 1167. & 865. & 642. & 464. & 194. & 39.\\
		\hline
		$log[n(z_j)] \propto$ & 4.69 & 3.67 & 3.19 & 3.07 & 2.94 & 2.81 & 2.67 & 2.29 & 1.59\\
		\hline
	\end{tabular}
\end{table*}

\begin{table*}
	\centering
	\caption{Results of earlier similar calculation for a sample of 76 quasars}
	\label{tab:Table 2}
	\begin{tabular}{lcccc} 
		\hline
		$z_m$-bin & 0 to 0.8 & 0.8 to 1.6 & 1.6 to 2.4 & 2.4 to 3.2\\
		\hline
		$z_j$(bin mid-pt) & 0.4 & 1.2 & 2.0 & 2.8\\
		\hline
		Bin pop. & 19 & 31 & 16 & 10\\
		\hline
		j (bin no.) & 1 & 2 & 3 & 4\\
		\hline
		$n(z_j)\propto$ & 1307. & 255. & 67. & 22.\\
		\hline
		$log[n(z_j)] \propto$ & 3.12 & 2.41 & 1.83 & 1.34\\
		\hline
	\end{tabular}
\end{table*}


\subsection{Differential distributions $p_i(x)$ of $x \equiv V/V_m$ for the nine $z_m$-bins}
For each of the 9 bins, indexed by $j = 1$ to 9, $p_i(x)$ histograms are plotted with $\Delta x = 0.2$ from $x = 0$ to 1, making five $x$-bins over the range [0, 1] of $x$. For $p_1(x)$, a curve is drawn by eye. For all other $p_i(x)$, $i = 2$ to 9, the extrapolated frequency polygon, with slightly higher slope than the last segment (to $x = 1$), is used. Cosmological number density $n(z_j)$ is then calculated from the formula (Banhatti 2009):

\begin{equation}
(\Omega/3)(c/H_0)^3 n(z_j) = \sum_{i=j}^9 (N_i /v(z_i))p_i(x_{ij}),
\end{equation}

where $x_{ij} = v(z_j)/v(z_i)$, and $\Omega$ is the survey solid angle. In this formula, $N_i$ are the bin populations of the 9 bins. Details of $n(z_j)$ calculation are shown in Table 3.
Examples are given below.

$n(z_7) \propto \sum_{i=7}^9 (N_i /v(z_i))p_i(x_{i7}) = (N_7 /v(z_7))p_7(x_{77}) + (N_8 /v(z_8))p_7(x_{87}) + (N_9 /v(z_9))p_7(x_{97})$.

The $p_7(x_{77})$, $p_7(x_{87})$ \& $p_7(x_{97})$ values are interpolated from the $p_7(x)$ frequency polygon.
Thus, for $n(z_1)$ calculation, there are 9 terms to sum (many of which happen to be 0 due to $p_1(x_{i1})$ being 0). For $n(z_2)$ there are 8 terms, and so on. Finally, for $n(z_9)$ there is only one term: $n(z_9) \propto (N_9 /v(z_9))p_9(x_{99}) = (31/2.112)2.65 = 38.9 \approx 39$.

\begin{table*}
	\centering
	\caption{Calculation of n(z(j)). Values in rows labelled i = 1, i = 2 \& so on are p(x(ij)).}
	\label{tab:Table 3}
	\begin{tabular}{lccccccccc} 
	\hline
	j & 1 & 2 & 3 & 4 & 5 & 6 & 7 & 8 & 9\\
	\hline
	$N_i$ & 29 & 34 & 34 & 32 & 33 & 32 & 30 & 31 & 31\\
	\hline
	$z_j$ & 0.15 & 0.50 & 0.95 & 1.35 & 1.65 & 2.00 & 2.50 & 3.40 & 300*\\
	\hline
	$v(z_j)\propto$ & 0.002231 & 0.03835 & 0.1251 & 0.2126 & 0.2773 & 0.3492 & 0.4436 & 0.5890 & 2.112\\
	\hline
	i = 1 & 1 & & & & & & & & \\
	\hline
	i = 2 & 0.058 & 1 & & & & & & & \\
	\hline
	i = 3 & 0.018 & 0.307 & 1 & & & & & & \\
	\hline
	i = 4 & 0.010 & 0.180 & 0.588 & 1 & & & & & \\
	\hline
	i = 5 & 0.008 & 0.138 & 0.451 & 0.767 & 1 & & & & \\
	\hline
	i = 6 & 0.006 & 0.110 & 0.358 & 0.609 & 0.794 & 1 & & & \\
	\hline
	i = 7 & 0.005 & 0.086 & 0.282 & 0.479 & 0.625 & 0.787 & 1 & & \\
	\hline
	i = 8 & 0.004 & 0.065 & 0.212 & 0.361 & 0.471 & 0.593 & 0.753 & 1 & \\
	\hline
	i = 9 & 0.001 & 0.018 & 0.059 & 0.101 & 0.131 & 0.165 & 0.210 & 0.279 & 1\\
	\hline
	$n(z_j) \propto$ & 48770. & 4717. & 1560. & 1167. & 865. & 642. & 464. & 194. & 39.\\
	\hline
	\end{tabular}
\end{table*}

\section{Results, Discussion and Conclusion}

Plots of numbers proportional to $n(z_j)$ and $log[n(z_j)$] listed in Tables 1 and 2 against $z_j$ show the following broad trends. The linear plot falls steeply towards large $z$ from a high value at the lowest $z$ (0.15 for the sample of 286 \& 0.4 for the 76). The log-linear plot brings out the variation at larger $z$ values more clearly. For the smaller sample of 76, a straight line of falling (i.e., negative) constant slope is a very good approximation. The log-linear plot for the larger sample of 286, which starts at a significantly lower $z$ value, falls more steeply than the smaller sample initially, and then the slope becomes shallower (less negative) than the smaller sample.
\\
Using $ln(z_m)$ in place of $z_m$ in the whole analysis leads to essentially the same results.
\\
The cosmological number density $n(z)$ is interpreted as the redshift distribution $n(z; S_0)$, which is the integral of the radio luminosity function $\rho(P, z)$ over all luminosities present in the sample as determined by the flux density limit. Thus,

\begin{equation}
n(z; S_0) = \int_{S_0 \ell_\nu^2(\alpha, z)}^\infty \rho(P, z)dP.
\end{equation}

This interpretation of $n(z)$ [derived from the differential distribution $p(x) \equiv p(V/V_m)$] as the redshift distribution  $n(z; S_0)$ needs to be explored further and utilized in deriving the cosmological evolution of the source population (here quasars).












\section*{Acknowledgements}

The work reported was done at School of Physics, Madurai Kamaraj University, Madurai, India. I thank Vasant Kulkarni for introducing me to $V/V_m$ test and luminosity functions.












\bsp	
\label{lastpage}
\end{document}